\documentclass[%
 reprint,
 superscriptaddress,
 amsmath,amssymb,
 aps,
 pre,
 showkeys,
]{revtex4-2}

\usepackage [english]{babel}
\usepackage{bm} 
\usepackage{enumitem} 
\usepackage{mathtools}
\usepackage{graphicx,caption} 
\usepackage [autostyle, english = american]{csquotes}
\MakeOuterQuote{"}

\DeclarePairedDelimiter\floor{\lfloor}{\rfloor}

\newcommand{\nodes}{N}
\newcommand{\Sperc}{S_\infty}
\newcommand{\Srec}{S_\mathrm{rec}}
\newcommand{\Savg}{\bar{S}}

\begin{document}
\title{Robustness of `small' networks}

\author{Jessica Jiang}
 \affiliation{Department of Mathematics, Dartmouth College, Hanover, New Hampshire 03755, USA.} 
\author{Allison C.\:Zhuang}
\affiliation{%
 Department of Mathematics, Dartmouth College, Hanover, New Hampshire 03755, USA.}%
\author{Petter Holme}
\affiliation{%
 Department of Computer Science, Aalto University, Espoo 02150, Finland.}%
\author{Peter J.\:Mucha}
\affiliation{%
Department of Mathematics, Dartmouth College, Hanover, New Hampshire 03755, USA.}%
\author{Alice C.\:Schwarze}%
\affiliation{%
 Department of Mathematics, Dartmouth College, Hanover, New Hampshire 03755, USA.}%
 
\date{\today}

\begin{abstract}
Modeling how networks change under structural perturbations can yield foundational insights into network robustness, which is critical in many real-world applications. The largest connected component is a popular measure of network performance. Percolation theory provides a theoretical framework to establish statistical properties of the largest connected component of large random graphs. However, this theoretical framework is typically only exact in the large-$\nodes$ limit, failing to capture the statistical properties of largest connected components in small networks, which many real-world networks are. We derive expected values for the largest connected component of small $G(\nodes,p)$ random graphs from which nodes are either removed uniformly at random or targeted by highest degree and compare these values with existing theory. We also visualize the performance of our expected values compared to existing theory for predicting the largest connected component of various real-world, small graphs.
\end{abstract}

\maketitle

\section{Introduction\label{sec:introduction}}

Real-world networks, such as power grids or the Internet, require a certain level of connectivity between the nodes to function. It is common to assess the performance and integrity of a network via the fraction $S$ of a network's nodes that are included in its largest connected component (LCC). For a network with $\nodes$ nodes and a largest connected component of $n$ nodes, we have
\begin{equation}
    S = \frac{n}{\nodes}.\label{eq:S}
\end{equation}
A relative LCC size close to 1 indicates a network in which most nodes can pass information and communicate with most of the other nodes through some route along edges. The relative LCC size is an important quantity for modeling and analyzing the robustness of a network \cite{Holme2002,Motter2004,Louzada2013}.

Our understanding of network robustness is furthered by percolation theory, a theoretical framework that, among other things, provides a calculation for the expected value of $S$ for various random-graph models in the infinite-$\nodes$ limit. For the Erd\H{o}s--R\'enyi (ER) $G(\nodes, p)$ random-graph model, the expected relative size of the LCC in the infinite-$\nodes$ limit, $\Sperc$, depends on the edge probability $p\in[0,1]$. The graph $G(\nodes, 0)$ of isolated nodes has $\Sperc=0$, and the complete $G(\nodes, 1)$ graph has $\Sperc=1$. Starting at $p=0$ and gradually increasing the edge probability, $\Sperc$ undergoes a phase transition from zero to positive values at a critical density denoted $p_c$. For $G(\nodes,p)$, this transition occurs at the  density where mean degree $\langle k\rangle$ satisfies $\langle k\rangle_c=(\nodes-1)p_c=1$. 
In the infinite-$\nodes$ limit, the expected LCC size, $\Sperc$, for $G(\nodes,p)$ obeys
\begin{equation}
    \Sperc = 1 - e^{-\langle k\rangle \Sperc}\,,
    \label{eq:relative LCC}
\end{equation}
which one can solve using the Lambert $W$ function \cite{Newman}.

\begin{figure}[t]
\centering
\includegraphics[scale=0.66]{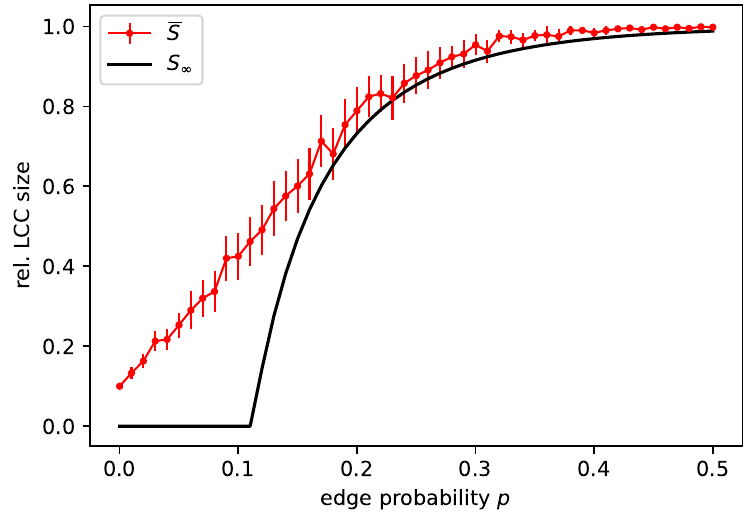}
\caption{{\bf Percolation in finite graphs.} Comparison of the expected LCC size from percolation theory, $\Sperc$, (black solid line) to the mean LCC size, $\Savg$, of 100 simulations of a $G(10,0.1)$ graph (red markers) with error bars indicating three standard errors of the $\Savg$ (which correspond to an approximate confidence interval of 99.7\%).}
\label{fig:Introfig} 
\end{figure}

While $\Sperc$ obtained from \eqref{eq:relative LCC} is very accurate for $\nodes$ sufficiently large, provided $p$ isn't too close to $p_c$,
it is well known that the range of $p$ for which finite-size effects make this approximation inaccurate expands dramatically for smaller networks. We demonstrate this inaccuracy in Fig.\,\ref{fig:Introfig}, where we plot $\Sperc$ for a $G(10,p)$ random graph and $\Savg$, the sample mean of $S$ from 100 realizations. The curve of $\Sperc$ indicates that the percolation threshold in the large-$\nodes$ limit is $p_c\approx0.11$. While the kink in $\Sperc$ at the percolation threshold $p_c$ clearly marks the phase transition between a random graph with a giant component ($\Sperc>0$) to one without ($\Sperc=0$), there is no similar indication for $\Savg$. 

We note that the percolation threshold of a graph can also be reached by varying $\nodes$, rather than by varying $p$ explicitly, as larger values of $\nodes$ will result in larger average degrees $\langle k \rangle = (\nodes - 1)p.$ Starting with an empty $G(0,p)$ graph and adding nodes will thus result in a similar curve to Fig.\,\ref{fig:Introfig}. This process will be highlighted in this paper.

The discrepancy between $\Sperc$ and $\Savg$ can lead to incorrect reference values when analyzing the robustness of networks \cite{Oehlers2021}. To address this discrepancy, we calculate the expected relative LCC size $\langle S\rangle$ for small finite-$\nodes$ graphs using combinatorial rather than statistical theory. In Section \ref{sec:theory}, we introduce our method for calculating the expected relative LCC size for small networks and its progression as one removes nodes from a network. In Section \ref{resultssec}, we compare our results to $\Sperc$ for various $G(\nodes,p)$ models. In Section \ref{real networks}, we compare our predictions to computational node-removal experiments on small real-world networks to demonstrate the effectiveness of our method in real-world applications. In Section \ref{conclusion}, we provide a final overview and conclusions of our findings. The code for our calculations and experiments is available online \cite{code}.

\section{Theory}\label{sec:theory}

We now introduce the theoretical framework that we propose for modeling the expected relative size $\langle S\rangle$ of the LCC of a small $G(\nodes,p)$ random graph and its change as one removes nodes from the graph. In Section \ref{sec:lccsize}, we derive a set of equations to calculate the expected size $\langle n \rangle$ of the LCC in a $G(\nodes,p)$ random graph of arbitrary size. In Section \ref{sec:lccsize}, we construct a model for the change of the relative size of the LCC in a $G(\nodes,p)$ random graph as one sequentially removes nodes from it. We consider choosing nodes to be removed either uniformly at random or targeted by node degree.

\subsection{Size of the largest connected component}\label{sec:lccsize}

For an ensemble of random graphs with $\nodes$ nodes, the expected size $\langle n \rangle$ of its LCC is, by definition,
\begin{equation}
\langle n \rangle = \sum_{n=1}^N nP_n\,,
\end{equation}
where $P_n$ is the probability that a graph in the ensemble has an LCC of size $n$. For the $G(N,p)$ random-graph model, $P_n$ and $\langle n \rangle$ depend on the model parameters $\nodes$ and $p$. We write
\begin{equation}
\langle n\rangle(\nodes,p) = \sum_{n=1}^\nodes nP_n(\nodes,p)\,.\label{eq:expLCC}
\end{equation}

A graph can have two or more LCCs, which are disjoint from each other and each include $n$ nodes for some $n\leq \nodes/2$. The probability of a graph having more than one LCC is small for large graphs and vanishes as $\nodes\to\infty$. For small graphs, however, the probability of the existence of multiple LCCs is much higher. For example, simple graphs with $\nodes=2$ nodes have just two possible configurations: the complete 2-node graph and the empty 2-node graph, the latter involving two LCCs of size 1. Simple graphs with $\nodes=4$ can have multiple LCCs with $n>1$: of the 64 possible configurations (distinct under node indexing), 6 configurations have two LCCs of size 2 (plus the empty graph configuration with four LCCs of size 1). These examples illustrate that a theory of percolation in small graphs should account for the occurrence of multiple LCCs.

There can be a maximum of $\floor{\frac{\nodes}{n}}$ components of size $n$. Denoting  the probability of a $G(\nodes,p)$ random graph having exactly $i$ largest connected components of size $n$ by $P_{n,i}(\nodes,p)$, we have that
\begin{equation}
    P_n(\nodes,p) = \sum_{i=1}^{\floor{\frac{\nodes}{n}}} P_{n,i}(\nodes,p).
    \label{eq:pn}
\end{equation}

The probability $P_{n,1}$ of a $G(\nodes,p)$ random graph having exactly one LCC and that LCC having size $n$ is
\begin{equation}
    P_{n,1}(\nodes,p) = {\nodes\choose n}\,h(n,p)\,g_n(\nodes,p) \sum_{j=1}^{n-1} P_{j}(\nodes-n,p)\,,\label{eq:pn1}
\end{equation}
where 
\begin{equation}
g_n(\nodes,p):=(1-p)^{n(\nodes-n)}
\label{eq:g}
\end{equation}
is the probability that a given set of $n$ vertices in a $G(\nodes,p)$ random graph is not connected to any of the other $\nodes-n$ vertices, and
\begin{equation}
h_n(p) := 1-\sum_{j=1}^{n-1}\binom{n-1}{j-1}\,h_j(p)\,g_j(n,p)
\end{equation}
is the probability that the induced subgraph on a given set of $n$ nodes in 
a $G(\nodes,p)$ random graph is connected. We explain \eqref{eq:pn1} as follows: There are ${\nodes\choose n}$ ways to pick a set of $n$ nodes in a graph of $\nodes$ nodes. The set of $n$ nodes forms an LCC, without any other LCC of matching size in the graph, if all three of the following conditions hold: 
\begin{enumerate}[itemsep=0mm]
    \item[1.] they are connected to each other,
    \item [2.]they are not connected to any of the remaining $\nodes-n$ nodes, and 
    \item[3.] the LCC in the remaining graph of $\nodes-n$ nodes is strictly smaller than $n$.
\end{enumerate} In a $G(\nodes,p)$ random graph, these three conditions are independent events, and the respective probabilities for conditions 1, 2, and 3 are $h_n(p)$, $g_n(\nodes,p)$, and $\sum_{i=1}^{n-1}P_i(\nodes-n,p)$.

To determine $P_{n,i}(\nodes,p)$ for $i>1$, we build the probability in a similar fashion. There are $\binom{\nodes}{n}$ ways to select the first set of $n$ nodes, $\binom{\nodes-n}{n}$ ways to subsequently choose the second set of $n$ nodes, $\binom{\nodes-2n}{n}$ ways to choose a third set of $n$ nodes, and so on. The number of ways to choose $i$ distinct sets of $n$ nodes is thus
\begin{equation}
\nu_{n,i}(\nodes)=\frac{1}{i!}\prod_{j=1}^i {\nodes-(j-1)n \choose n}\,,\label{eq:numsets}
\end{equation}
where the denominator accounts for the permutations of the order of the $i$ sets of $n$ nodes.
The selected sets form the $i$ LCCs of size $n$ if 
\begin{enumerate}[itemsep=0mm]
\item[1'.] each set is a set of connected nodes,\vspace{-3mm}
\item[2'.] each set is not connected to any of the remaining nodes, and \vspace{-3mm}
\item[3'.] the LCC in the remaining graph of $\nodes-in$ nodes is strictly smaller than $n$.
\end{enumerate}
 In a $G(\nodes,p)$ random graph, the probability of condition 1' to be fulfilled is $h_n(p)^i$, and the probability for condition 2' to be fulfilled is
\begin{equation}
\gamma_{n,i}(\nodes,p)=\prod_{j=1}^i g_n(\nodes-(j-1)n,p)\,.
\end{equation}
The probability that condition 3' is fulfilled is 
\begin{equation}
    \pi_{n,i}(\nodes,p) = \sum_{j=1}^{n-1} P_j(\nodes-in,p)\,.
\end{equation}
We thus have
\begin{equation}
    P_{n,i}(\nodes,p)
    = \nu_{n,i}(\nodes)\,[h_n(p)]^i\,\gamma_{n,i}(\nodes,p)\,\pi_{n,i}(\nodes,p)
    \,.\label{eq:pni}
\end{equation}

Equations \eqref{eq:expLCC}--\eqref{eq:pni} together yield a theoretical framework for calculating the expected size $\langle n\rangle$ of the LCC in a $G(\nodes,p)$ random graph via a recursive scheme starting from the probabilities $P_{n=1}(\nodes', p)$ for $\nodes'\in\{1,\dots,\nodes\}$. Probability $P_{n=1}(\nodes',p)$ is the probability that a $G(\nodes',p)$ random graph has an LCC of size 1, which is true if and only if the graph has no edges. Thus, the initial condition for our recursive scheme is
\begin{equation}
    P_{n=1}(\nodes,p) = (1-p)^{\nodes \choose 2}.\label{eq:initial}
\end{equation}
We use the notation $\Srec$ to specially denote the expected relative LCC size obtained by the recursive scheme described in equations \eqref{eq:expLCC}--\eqref{eq:initial}, with $\Srec = \frac {\langle n \rangle } {N}$.

\subsection{Impact of node removal}

We consider the disintegration of networks under sequential node removal, which is a common approach for analyzing the structural robustness of networks \cite{Sahimi2023}.
In this scenario, the expected relative LCC size is a function $\langle S\rangle(r)$ of the number $r$ of nodes that have been removed from the original graph. The removal of $r$ nodes from a $G(\nodes,p)$ graph creates an ensemble of random graphs of the remaining $\nodes-r$ nodes. The properties of this random-graph ensemble depend on how nodes are removed from the original graph, and it is not necessarily equivalent to the $G(\nodes-r,p)$ random-graph model. 

Representing a graph $G$ by its adjacency matrix $A$, a random-graph model is a probability distribution $P(A)$ of adjacency matrices. When removing a node $i_R$ from a graph $G$ with $\nodes$ nodes, the resulting graph $G'$ has an $(\nodes-1)\times(\nodes-1)$ adjacency matrix $A'$. The joint probability of $A'$ and the $i_R$th row, $A[i_R]$, of $A$ is equivalent to $P(A)$. The conditional probability $P(A'\,\vert\,i_R)$ of adjacency matrices $A'$ given the removal of node $i_R$ from $G$ is thus the marginal distribution
\begin{equation}
P(A'\,\vert\,i_R)=\sum_{A[i_R]}P(A\,\vert\,i_R)P(A[i_R]\,\vert\,i_R)\,.
\end{equation}
The conditional probability $P(A\,\vert\,i_R)$ is connected to the node removal process via Bayes' theorem,
\begin{equation}
P(A\,\vert\,i_R)=\frac{P(i_R\,\vert\,A)P(A)}{P(i_R)}\,,\label{eq:bayes}
\end{equation}
where $P(i_R\,\vert\,A)$ is the probability of removing node $i_R$ from a network with 
adjacency matrix $A$. 

In the following subsections, we consider two node-removal processes: 1.~removing nodes uniformly at random, and 2.~removing nodes targeted by highest degree. 

\subsubsection{Random node removal}\label{sec:randomremoval}

The $G(\nodes,p)$ random-graph model corresponds to a binomial distribution with binomial probability $p$,
\begin{equation}
P(A)=\prod_{\substack{i=1,\\j>i}}^\nodes p^{a_{ij}}(1-p)^{1-a_{ij}}\,.
\end{equation}
When removing nodes uniformly at random, the node selection is independent of the network's structure (i.e., $P(i_R|A)=P(i_R)=1/\nodes$) and $A'$ and $A[i_R]$ are independent random variables (i.e., $P(A',A[i_R])=P(A')P(A[i_R])$) with binomial distributions with binomial probability $p$. The resulting probability distribution after a first node removal is the binomial distribution 
\begin{equation}
    P(A'|i_R)=\prod_{\substack{i=1,\\ j>i}}^{\nodes-1} p^{a_{ij}'}(1-p)^{1-a_{ij}'}\,,
\end{equation}
which is the probability distribution of the $G(\nodes-1,p)$ random graph.
Removing one node uniformly at random from the $G(\nodes,p)$ random graph thus leads to the $G(\nodes-1,p)$ random graph. Further successive node removals until $r$ nodes have been removed then lead to the $G(\nodes-r,p)$ random graph, provided all removed nodes have been chosen uniformly at random. The expected relative size of the LCC of a network from which $r$ nodes have been removed uniformly at random is thus equal to $\langle S \rangle$ of the $G(\nodes-r,p)$ random graph, and it can be calculated as $\Srec$ via \eqref{eq:expLCC}--\eqref{eq:initial}.

\subsubsection{Degree-targeted node removal}
\label{section:attacks}

In the case of degree-targeted removals, the highest-degree node is removed at each perturbation. When removing the highest-degree node, the node selection is not independent of the network's structure (i.e., $P(i_R|A)\neq P(i_R)$). The removal of the highest-degree node from the $G(\nodes,p)$ random graph thus results in a new random-graph ensemble that is different from $G(\nodes-1,p)$. We denote this new random-graph ensemble of $N-1$-sized graphs $G'(\nodes,p)$, explicitly indicating the size $\nodes$ and edge probability $p$ of the original ER graph. Notably, the new ensemble $G'(\nodes,p)$ has edge density lower than $p$, and a degree distribution that is not binomial. We explore the degree distributions of subgraphs obtained via degree-targeted node removal and the relative LCC size of a $G(n,p)$ random graph under degree-targeted node removal in Section \ref{section:attacksresults}.

For the study of the expected relative LCC size $\langle S\rangle$ of a $G(\nodes,p)$ random graph under degree-targeted node removal, we make two simplifying assumptions:
\begin{enumerate}
    \item[A1.] The expected degree distributions of the subgraph ensemble $G'(\nodes, p)$ that one obtains via degree-targeted removal of any number of nodes from a $G(\nodes, p)$ random graph is a binomial distribution.
    \item[A2.] The degrees of nodes in a $G'(\nodes, p)$ random graph are independent binomial random variables.
\end{enumerate}

Employing A1, we approximate $G'(\nodes,p)$ by $G(\nodes,p')$, calculating $p'$ from the expected edge count $\langle m' \rangle$ of the $G'(\nodes,p)$ ensemble, which depends on the expected edge count $\langle m \rangle$ and the expected maximum degree $\langle k_{\max}\rangle$ of the $G(\nodes,p)$ random graph according to
\begin{align}
    p'=\frac{\langle m'\rangle}{{\nodes-1\choose2}} 
    = \frac{\langle m\rangle - \langle k_{\max}\rangle}{{\nodes-1\choose2}} 
    = \frac{{\nodes\choose2}p - \langle k_{\max}\rangle}{{\nodes-1\choose2}}\,.
    \label{eq:p'}
\end{align}
To calculate $p'$, the expected maximum degree is
\begin{equation}
    \langle k_{\max}\rangle = \sum_{i=0}^{\nodes-1} iP(k_{\max}= i)\,,
\end{equation}
for which we obtain the probability distribution of the maximum degree via differencing:
\begin{equation}
    P(k_{\max}= i) = P(k_{\max}\geq i) - P(k_{\max}\geq i+1)\,.
\end{equation}
Under A2, the cumulative probability
\begin{equation}
    P(k_{\max}\geq i) = 1-(1-P(k\geq i))^\nodes\,,
\end{equation}
where the probability $P(k\geq i)$ of a chosen node in the graph to have degree $k\geq i$ follows the cumulative binomial distribution
\begin{equation}
    P(k\geq i) = \sum_{j=i}^{\nodes-1}{\nodes-1 \choose j}p^j(1-p)^{\nodes-1-j}\,.
\end{equation}

We can then use the $p'$ resulting from the above assumptions as input to our recursive scheme calculation for $\Srec$ and the large-$\nodes$ percolation prediction $\Sperc$. Because of the inaccuracies in the above assumptions, however, we do not expect this application of $\Srec$ for targeted node removal to precisely equal the true $\langle S\rangle$.

\section{Robustness of small random graphs} \label{resultssec}

In this section, we explore how accurately our recursive scheme $\Srec$ captures the relative LCC size of realizations of a $G(\nodes,p)$ model upon node removals. In subsection \ref{sec:res1}, we compare the relative LCC size under uniform-at-random and degree-targeted node removal for a $G(\nodes,p)$ with fixed $\nodes$ and $p$. In subsection \ref{section:fig2A}, we explore how the relative LCC size under uniform-at-random node removal depends on $\nodes$ and $p$. We present results on the parameter-dependence of the relative LCC size under degree-targeted node removal in subsection \ref{section:attacksresults}. In subsection \ref{sec:res4}, we compare the accuracy of our recursive scheme with that of the large-$\nodes$ limit $\Sperc$ from percolation theory.

\subsection{The relative size of the largest connected component of small $G(\nodes,p)$ random graphs}\label{sec:res1}

\begin{figure}[t]
\centering
\includegraphics[scale=0.55]{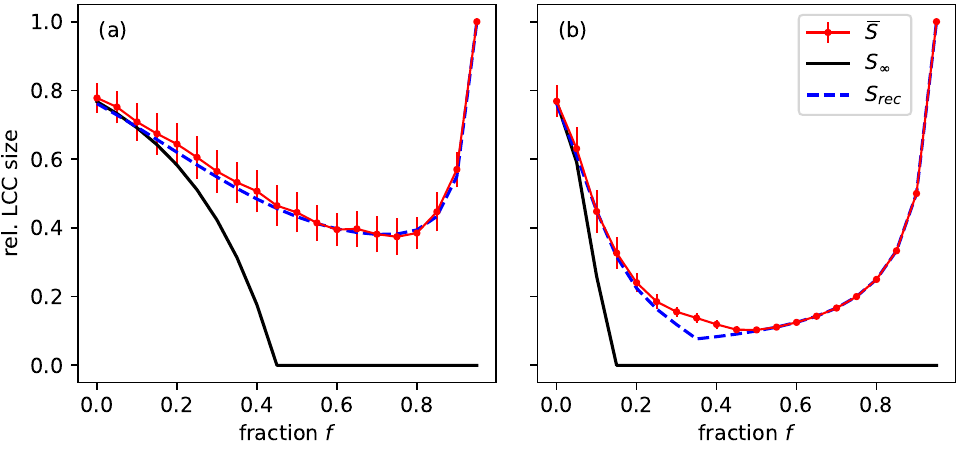}
\caption{{\bf Uniform-at-random versus degree-targeted node removal.} Relative LCC sizes for $G(20,0.1)$ random graphs after removing proportion $f$ of the edges of each graph by (a) uniform-at-random removals and (b) degree-targeted removals, comparing the sample mean $\Savg$ from 100 realizations (error bars indicate 3 standard errors), the sizes $\Srec$ predicted by our recursive scheme, and those predicted by large-$\nodes$ percolation theory, $\Sperc$.}
\label{fig:Fig1}
\end{figure}

\begin{figure*}[t]
\centering
\includegraphics[width=1\textwidth]{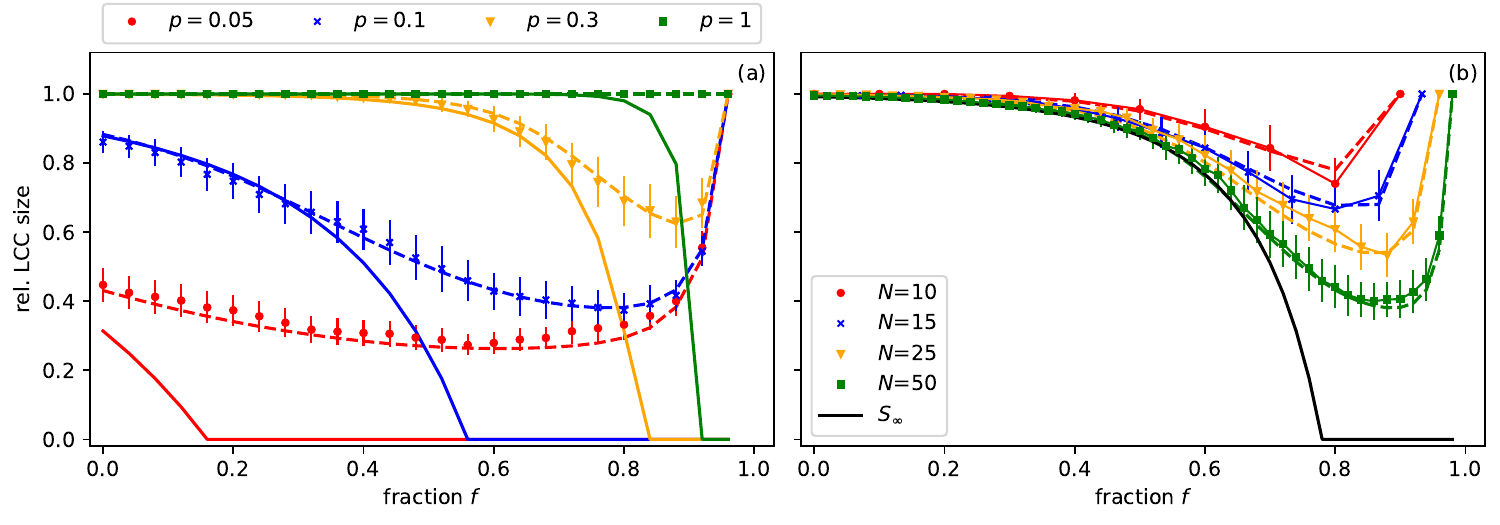}
\caption{{\bf The relative size of the largest connected component under uniform-at-random node removal.} Markers show simulated results $\Savg$ of the relative LCC size as functions of the fraction of nodes $f$ removed from $G(\nodes,p)$ random graphs through uniform-at-random removals. In panel (a), $\nodes = 25$. The dashed and solid lines of the same color show the predictions $\Srec$ and $\Sperc$, respectively. In panel (b), for each initial $\nodes$ the edge density $p$ is set by $(\frac{\nodes}{5}-1)p=1$, rounded to two decimal places. Dashed lines of the same color again show the prediction $\Srec$, but $\Sperc$ is provided as a solid black line because it is the same for every $(\nodes,p)$ combination satisfying $(\frac{\nodes}{5}-1)p=1$.}
\label{fig:Fig2}
\end{figure*}

As a first check of the accuracy of our model for LCC size of random graphs under node removal, we consider $G(\nodes,p)$ graphs with $\nodes=20$ and $p=0.1$. In Fig.\,\ref{fig:Fig1}, we show the (sample) mean relative LCC sizes $\Savg$ after removing proportion $f$ of the nodes from each of 100 realizations of $G(20,0.1)$, compared with the expected relative LCC size according to our recursive scheme, $\Srec$, and the expected relative LCC size according to percolation in the large-$\nodes$ limit, $\Sperc$. In panel (a), we show how these quantities change as one removes nodes uniformly at random. We show results for degree-targeted removal in panel (b). In both panels, error bars indicate three standard deviations of the distribution of $\Savg$. We select this convention to create visually detectable error bars, noting that an interval of three standard errors gives an approximate confidence interval of 99.7\%, indicating that the true relative LCC sizes would be expected to lie within this interval with almost full confidence.
 
When removing nodes uniformly at random (see Fig.\,\ref{fig:Fig1}(a)), the observed mean LCC size $\Savg$ initially decreases as the fraction of removed nodes, $f:=r/\nodes$, increases. Eventually, however, as the number of remaining nodes becomes small enough, this trend reverses as the absolute size $n$ of the largest connected component of the graph approaches its minimum value (i.e., $n_{\textrm{min}}=1$) and a continued removal of nodes affects the denominator in the calculation of the relative LCC size (see Eq.\,(\ref{eq:S})) more strongly than the numerator. Once one has removed all but one node from a graph, the remaining graph always has $S=1$.

The observed $\Savg$ under uniform-at-random node removal is captured well by the expected relative LCC size $\Srec$ obtained from our recursive scheme: for each value of $f$, $\Srec$ falls within three standard errors of $\Savg$. In contrast, the expected LCC size $\Sperc$ from percolation in the large-$\nodes$ limit deviates substantially from $\Savg$ and has a qualitatively different behavior. We can explain these differences as follows. Uniform-at-random node removal decreases a network's size, but leaves its expected edge density unaffected. The resulting decrease of $\langle k\rangle$ eventually falls below the percolation threshold for the infinite-$\nodes$ limit, $\langle k\rangle_c=(\nodes-1)p_c=1$. Consequently, the expected LCC size $\Sperc$ initially decreases with increasing $f$ (which decreases the remaining $\nodes$), drops to $0$ once $p_c=1/(\nodes-1)$ exceeds $p$, and remains 0 as even more nodes are removed. In contrast, the observed $\Savg$ is necessarily positive, deviates from $\Sperc$ as $f$ is pushed towards its infinite-$\nodes$ critical point, and has its largest difference with $\Sperc$ when all but the last node are removed from the network; that is, for the remaining single node, we have $\Savg=1$ while $\Sperc=0$.

When one targets nodes by degree (see Fig.\,\ref{fig:Fig1}(b)), the mean relative LCC size tends to decrease faster with increasing $f$ compared to uniform-at-random node removal, because of the removal of larger numbers of edges, and the trend reversal occurs at a smaller value of $f$. The targeted removal of nodes is also associated with a smaller variation in LCC sizes among realizations of the $G(\nodes,p)$ model, leading to shorter error bars on $\Savg$ in Figure \ref{fig:Fig1}(b). The expected relative LCC sizes $\Srec$ and $\Sperc$ obtained from our recursive scheme and percolation theory in the large-$\nodes$ limit also indicate a faster decline of the LCC under degree-targeted node removal than under uniform-at-random node removal. Similar to our observations on uniform-at-random node removal, we find that the expected LCC from our recursive scheme captures $\Savg$ much better than the percolation-theory results, though we note that $\Savg$ does not always fall within three standard errors of $\Savg$ for targeted node removal. 

\subsection{Uniform-at-random node removal from $G(\nodes,p)$ at different $\nodes$ and $p$}
\label{section:fig2A}

\begin{figure*}[t!]
\centering
\includegraphics[width=1\textwidth]{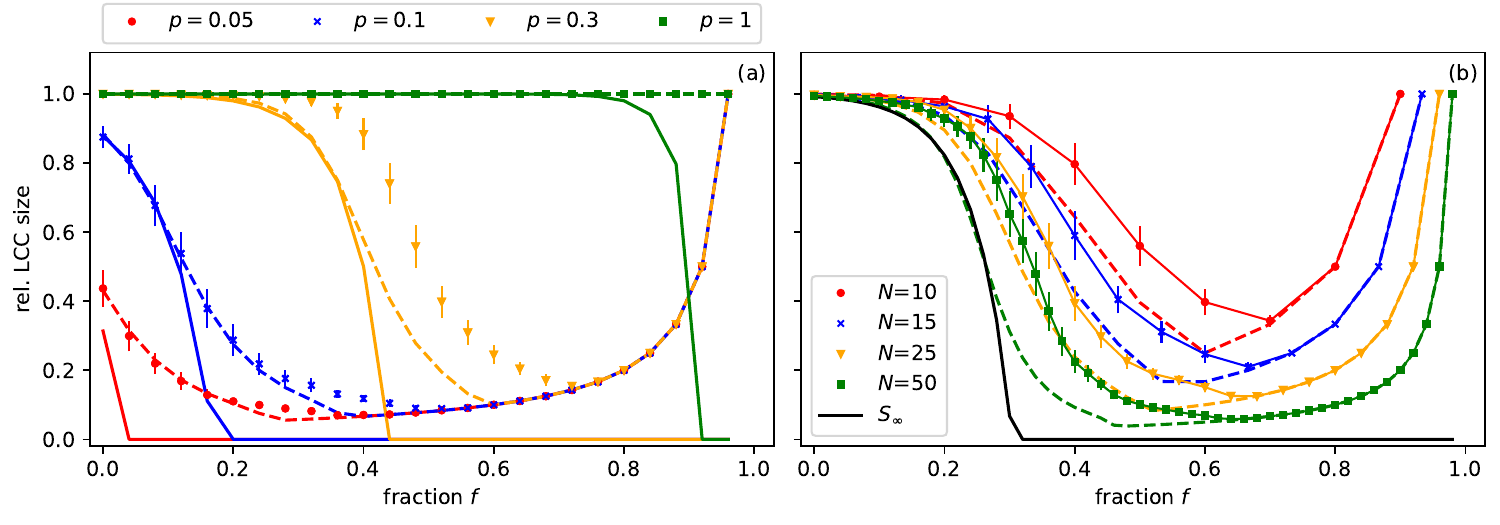} 
\caption{{\bf The relative size of the largest connected component under degree-targeted node removal.} Markers show simulated results $\Savg$ of the relative LCC size as functions of the fraction of nodes $f$ removed from $G(\nodes,p)$ random graphs through degree-targeted node removals. In panel (a), $\nodes = 25$. The dashed and solid lines of the same color show the predictions $\Srec$ and $\Sperc$, respectively. In panel (b), for each initial $\nodes$ the edge density $p$ is set by $(\frac{\nodes}{5}-1)p=1$, rounded to two decimal places. Dashed lines of the same color again show the prediction $\Srec$, but $\Sperc$ is provided as a solid black line because it is the same for every $(\nodes,p)$ combination satisfying $(\frac{\nodes}{5}-1)p=1$.}
\label{fig:Fig2-2}
\end{figure*}

Following our above exploration of $\Savg$, $\Srec$, and $\Sperc$ for an example $G(\nodes,p)$ random graph, we now consider how our observations are impacted by changes in $\nodes$ and $p$, with results plotted in Fig.\,\ref{fig:Fig2}. Varying the edge probability $p$ in Fig.~\ref{fig:Fig2}(a) with fixed $\nodes=25$, we observe that the trend reversal occurs at a smaller fraction of removed nodes for sparser random graphs (i.e., smaller $p$) than for denser random graphs (i.e., larger $p$). The corresponding minimum expected relative LCC size is larger for dense random graphs than for sparse random graphs.

The expected relative LCC size calculated via our recursive scheme, $\Srec$, accurately captures $\Savg$ for all considered values of $p$. In contrast, the expected relative LCC size calculated from percolation in the large-$\nodes$ limit, $\Sperc$, consistently fails to capture $\Savg$ for large $f$. Additionally, for any given value of $f$, the discrepancy between $\Sperc$ and $\Savg$ is larger for sparser random graphs (i.e., smaller $p$) than for denser ones.

In Fig.~\ref{fig:Fig2}(b), we tested our method for $G(\nodes,p)$ random graphs of varying sizes $\nodes$, setting the edge probability for each $\nodes$ so that the mean degree after removing $f=0.8$ of the nodes, i.e., when the number of remaining nodes is $\nodes'=\nodes/5$, satisfies $\langle k\rangle=p_c(\nodes'-1)p=0.2(\nodes-1)p=1$. Since uniform-at-random node removal does not change the expected edge density, the infinite-$\nodes$ percolation prediction $\Sperc$, which is identical for all $(\nodes,p)$ combinations in Fig.~\ref{fig:Fig2}(b), drops to 0 at $f=0.8$. Consistent with our previous observations above, the mean LCC size $\Savg$ of realizations of node removal from random graphs reduces substantially as $f$ approaches the critical point for percolation theory (i.e., in this case as $f$ approaches $0.8$) but this decrease is not as large as that predicted by $\Sperc$. The comparison of $\Savg$ and $\Sperc$ illustrates that the large-$\nodes$ limit percolation theory does not accurately capture the relative LCC size of random graphs as $f$ gets closer to the critical transition, with the deviation between $\Savg$ and $\Sperc$ happening at smaller $f$ for smaller $\nodes$. Moreover, and as we now expect, $\Sperc$ completely fails to explain $\Savg$ for $f$ above the transition. 

In contrast, the expected LCC size obtained via our recursive scheme, $\Srec$, captures the evolution of $\Savg$ under successive uniform-at-random node removal with high accuracy. In particular, for $\nodes=10$, 15, 25, 50, and 100 nodes, plotted in Fig.~\ref{fig:Fig2}(b), we obtain respective mean squared errors (MSEs) of 0.00038, 0.00096, 0.00052, $6.62 \times 10^{-5}$ and  $4.14 \times 10^{-5}$ when using $\Srec$ to predict $\Savg$. 

The use case for the large-$\nodes$ limit prediction, $\Sperc$, are large networks well before the fraction of removed nodes reaches the phase transition. The random graph $G(100,0.05)$ with $f\in[0,0.5]$ (not shown in Fig.\,\ref{fig:Fig2}) is an illustrative example; The MSE for fitting $\Savg$ with $\Sperc$ on this interval of $f$ is $1.46 \times 10^{-6}$, which is on par with the MSE of $1.53 \times 10^{-6}$ for fitting $\Savg$ with $\Srec$ on the same interval of $f$. 

\subsection{Degree-targeted node removal from $G(\nodes,p)$ random graphs at different $\nodes$ and $p$}
\label{section:attacksresults}

\begin{figure*}[t]
\centering
\includegraphics[width=1\textwidth]{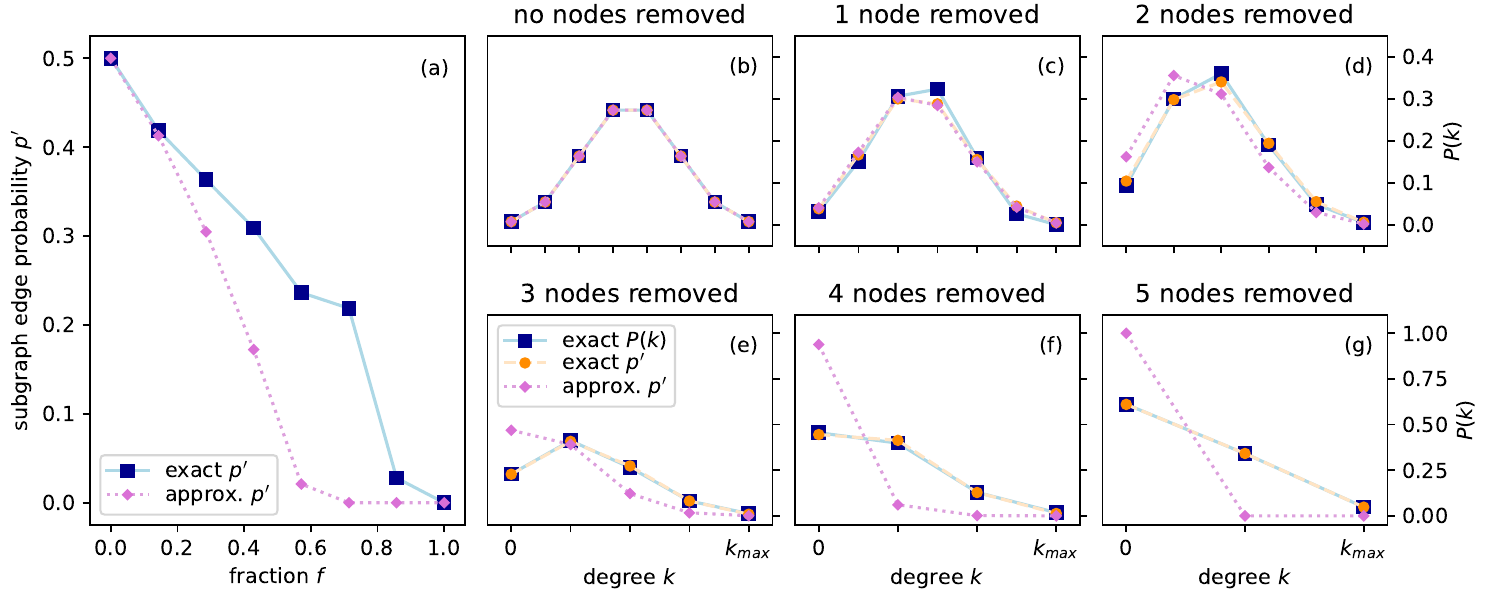} 
\caption{{\bf Edge density and degree distributions of subgraphs of $G(\nodes,p)$ random graphs under degree-targeted node removal.} In panel (a), we show the exact and approximated edge density of subgraphs of $G(\nodes,p)$ random graph with $\nodes=8$ and $p=0.5$ as a function of the fraction $f$ of nodes removed. In panels (b)--(g), we show the exact degree distribution as well as binomial distributions with exact and approximate edge densities throughout the process of removing 5 nodes targeted by degree.}
\label{fig:assumptions}
\end{figure*}

In Fig.\,\ref{fig:Fig2-2}, we compare $\Srec$ and $\Sperc$ for random graphs under degree-targeted node removal to the corresponding mean values $\Savg$ calculated from computational simulations. The expected LCC size $\Srec$ obtained via our recursive scheme more accurately captures $\Savg$ than the expected LCC size in the large-$\nodes$ limit, $\Sperc$, for a variety of edge probabilities $p$ and network sizes $\nodes$. However, as expected, we find that our approach is less accurate when used to predict $\Savg$ for degree-targeted node removals than for uniform random removals (cf.\ Fig.\,\ref{fig:Fig2}). The results in Fig.\,\ref{fig:Fig2-2} indicate that $\Srec$ provides a good fit for $\Savg$ for networks that are sufficiently sparse --- see, e.g., $p=0.05$ in Fig.\,\ref{fig:Fig2-2}(a). But the $\Srec$ prediction tends to underestimate the size of the LCC for moderate fractions $f$ of nodes removed at higher edge densities --- see $p=0.3$ in panel (a). The discrepancy between $\Srec$ and $\Savg$ starts to become apparent for $f$ below the critical fraction predicted by the percolation theory, and then continues up until the simulated mean size begins to increase with increasing $f$. This discrepancy might stem from either or both of the simplifying assumptions in our derivation (see Section \ref{section:attacks}) about 1.~the degree distribution of subgraphs of a random graph and 2.~the independence of degrees of nodes in a subgraph. In Fig.\,\ref{fig:assumptions}, we demonstrate a situation where the discrepancy specifically stems from A2, as described below.

Panel (a) of Fig.\,\ref{fig:assumptions} shows the edge probabilities $p'$ of subgraphs $G'(8,0.5)$ created through successive node removals from a $G(8,0.5)$ random graph while targeting nodes by degree. We obtain the exact edge probabilities $p'$ via exhaustive enumeration of all realizations of the $G(8, 0.5)$ random graph and its subgraphs. These exact values of $p'$ are compared with their approximations given by Eq.~(\ref{eq:p'}), which rely on both assumptions 1 and 2. As can be seen in panel (a), our approximation tends to underestimate the true edge probability. When our approximation of $p'$ substantially underestimates the true edge probability, the calculations of $\Srec$ and $\Sperc$ assume that the resulting subgraphs become sparser faster under degree-targeted node removals than they actually do. Accordingly, $\Srec$ and $\Sperc$ underestimate $\Savg$ for $f\in[0,0.5]$ in Fig.~\ref{fig:Fig2-2}.

In panels (b)--(g), we compare the exact degree distributions $P(k)$ of graph ensembles $G'(8,0.5)$ obtained via degree-targeted removal of 0 to 5 nodes from $G(8,0.5)$ random graphs to binomial degree distributions with the exact edge probabilities $p'$ and binomial degree distributions with our approximation of $p'$. We can see from panel (b) that when no nodes are removed, all three distributions align, as required. As nodes are removed, successively through panels (c)--(g)), the binomial distribution with our approximate $p'$ overestimates the probability of a node having a low degree. In contrast, the binomial distribution with the exact edge probability $p'$ aligns well with the exact $P(k)$ degree distribution. These observations indicate that our approximation error for degree-targeted node removal stems from our A2 (i.e., the assumption that the degrees of nodes in a small graph are independent random variables) rather than A1 (i.e., the assumption that $G'(\nodes,p)$ has a binomial degree distribution). Indeed, the degrees of nodes in a network are not independent of each other, 
and this lack of independence is especially prominent in small networks where each edge greatly affects the density of the network and thus the degree distribution. 

\begin{figure*}[t]
\centering
\includegraphics[scale=0.55, trim=6mm 0 6mm 0]{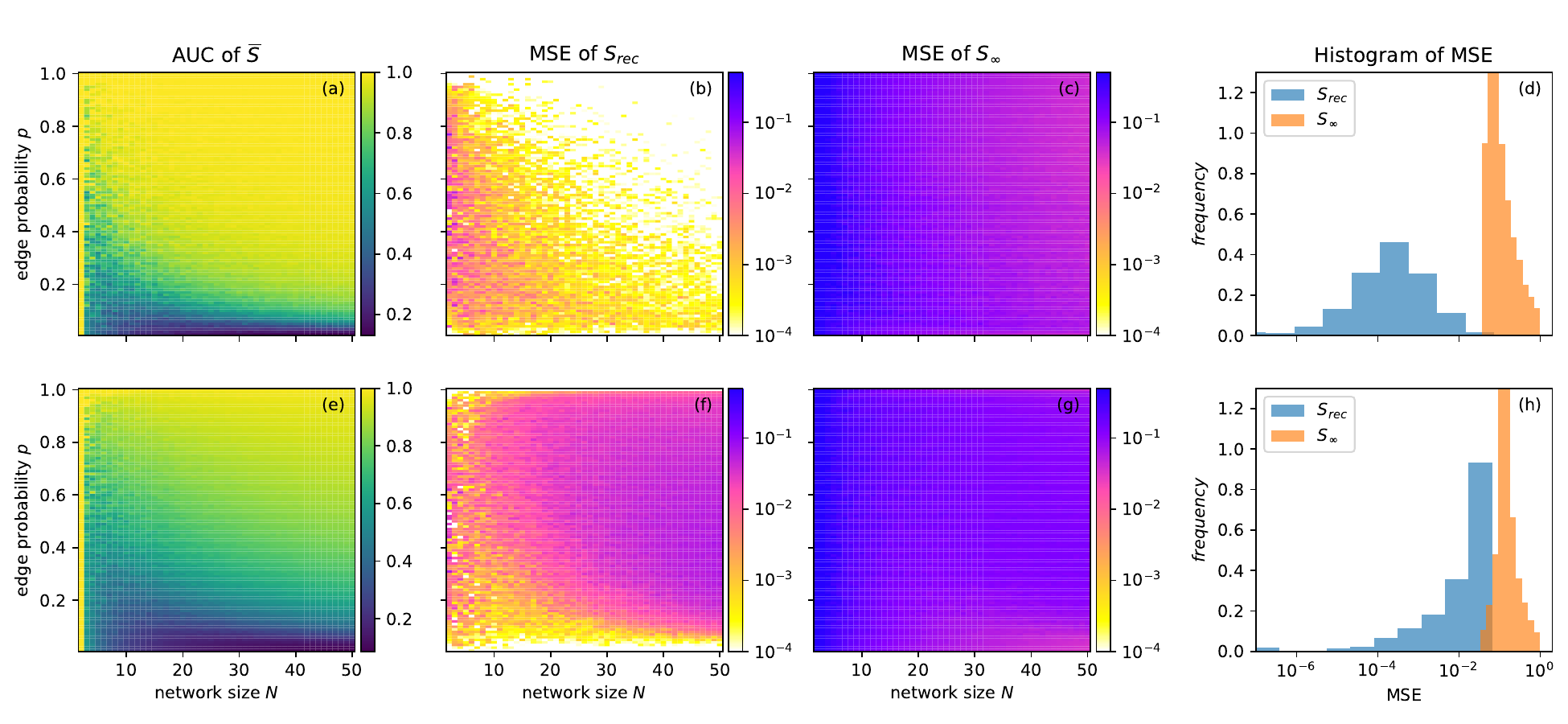}
\caption{{\bf Summary of results about relative LCC size upon node removal.} We show results across $p\in (0,1]$ for $\nodes\leq 50$. The top row, panels (a)--(d), shows results for uniform-at-random node removal, while the bottom row, panels (e)--(h), contains results for degree-targeted removal. Panels (a) and (e) show the area under the curve (AUC) of the simulated $\Savg$ across the full range of fraction of nodes removed, $f\in [0,1]$. The second and third columns show the mean squared error (MSE) of $\Srec$ (panels (b) and (f)) and $\Sperc$ (panels (c) and (g)) for predicting $\Savg$, plotted with a log-scaled heat map. Panels (d) and (h) summarize the results shown in panels (b), (c), (f) and (g) in  histograms.} 
\label{fig:Fig3}
\end{figure*}

\subsection{Dependence on parameters of the $G(\nodes,p)$ random graph model}
\label{sec:res4}

To give an overview of how well $\Srec$ and $\Sperc$ capture the decline of the LCC under uniform-at-random and degree-targeted node removal throughout the parameter space of the $G(\nodes,p)$ random graph model, we calculate the area under the curve (AUC) of $\Savg$ for the full range of the fraction of nodes removed, $f\in [0,1]$, and the mean squared errors (MSE) of using $\Srec$ and $\Sperc$ to fit $\Savg$, computed across the $\nodes$--$p$ plane for $\nodes\leq 50$.
The AUC of the LCC size over the fraction of removed nodes is a common measure of network robustness \cite{schwarze2024}, taking values in $[0,1]$. An AUC closer to 1 indicates that $S$ remains close to 1 for most of the node removal process, thereby signifying greater structural robustness of a network.

In Fig.\,\ref{fig:Fig3}(a), we see that $G(\nodes,p)$ random graphs tend to be most robust to uniform-at-random node removal when they are dense (i.e., they have a high edge probability $p$). Among random graphs with intermediate edge densities, larger graphs tend to be more robust than smaller graphs. These observations are consistent with our findings in Sections \ref{sec:res1} and \ref{section:fig2A}. Focusing on the differences between Fig.\,\ref{fig:Fig3} panels (a) and (e), one can see that random graphs undergoing uniform-at-random node removals tend to have a larger AUC than random graphs undergoing degree-targeted node removals, consistent with reports that $G(\nodes,p)$ random graphs tend to be robust to random failures but more sensitive to degree-targeted attacks \cite{Barabasi-Jeong-Albert, schwarze2024}.

Comparing the heatmaps in Fig.\,\ref{fig:Fig3}(b) and (c), it is evident that $\Srec$ calculated via our recursive scheme captures $\Savg$ under uniform-at-random node removal with much greater accuracy than $\Sperc$ for $G(\nodes,p)$ random graphs of at least up to 50 nodes (and presumably well past $\nodes=50$, given the observed trends in these panels). One can make similar observations for degree-targeted node removal from panels (f) and (g), with the errors becoming increasingly comparable as $\nodes$ increases towards $\nodes=50$ in the panels. Due to the limitations of our approximate calculation of edge probabilities $p'$ of subgraphs of $G(\nodes,p)$ graphs obtained via degree-targeted node removals (see Section \ref{section:attacksresults}), the MSEs for all considered parameter combinations tend to be much larger for degree-targeted node removal than for uniform-at-random node removal.

The MSE values shown in the heatmaps in Fig.~\ref{fig:Fig3}(b) and (c) are summarized in a histogram in Fig.~\ref{fig:Fig3}(d). (Similarly, the data from panels (f) and (g) are presented in histogram form in (h).) These histograms underscore our previous observations that the results of our recursive scheme capture $\Savg$ with greater accuracy than $\Sperc$ for all considered combinations of $\nodes$ and $p$, and that they yield more accurate results for uniform-at-random node removal than for degree-targeted node removal.

\begin{figure*}[t]
\centering
\includegraphics[scale=0.55, trim=10mm 0 5mm 5mm]{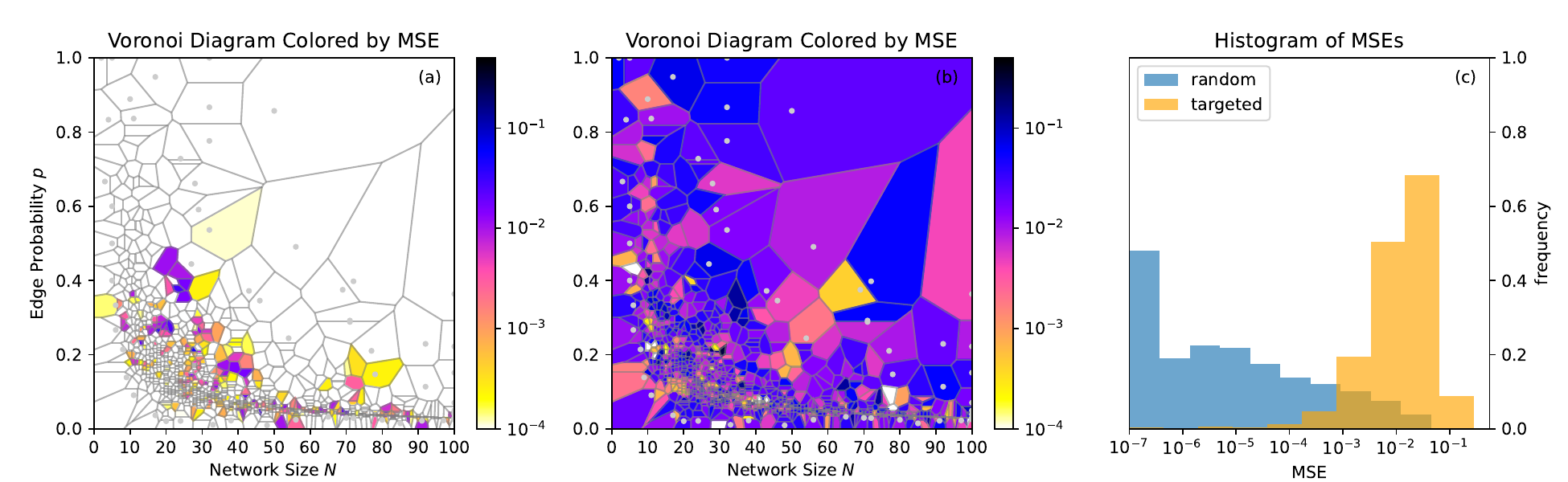}
\caption{{\bf Predicting robustness of real-world networks.} We show a Voronoi diagrams of the mean squared error (MSE) between our $\Srec$ predictions and the sample mean $\Savg$ for various real-world networks undergoing node removal. In panel (a), we show results for uniform-at-random node removal. In panel (b), we show results for degree-targeted node removal. MSE values below $10^{-4}$ are indicated as white cells. To increase the visibility of small Voronoi cells, their centers are not shown. The histograms in panel (c) summarize the MSEs for networks undergoing (blue) uniform-at-random node removal and (orange) for degree-targeted node removal. For visualization simplicity, we set MSE values below $10^{-7}$ equal to $10^{-7}$. Additionally, we removed three networks with large errors due to their extreme sparsity from the distribution.}
\label{fig:Fig4}
\end{figure*}

\section{Robustness of small real-world networks}
\label{real networks}

Our model typically results in small errors for predicting $\Savg$ for $G(\nodes,p)$ graphs, especially those undergoing random node removals. Real-world networks, however, are typically not well described by random graphs \cite{vanDerHofstadt2017}. We considered 1562 real-world networks obtained by gathering everything we could find that could be represented by a graph of $\leq 100$ nodes. The word ``real-world'' should thus be interpreted as generously as possible because we include networks invented by humans---to prove theorems in combinatorial game theory, for benchmarking algorithms, and works of art. The largest data sets come from animal social networks and transition graphs in bird song (and other forms of animal vocalization). For further discussion of the data and references, see \url{https://github.com/pholme/small}.

To test the accuracy of our method for many small real-world networks, we first simulated the random node removals on the real-world networks. We performed each node removal process on each network 100 times to calculate the sample mean $\Savg$ as nodes were removed. To apply our combinatorial calculations, based on $G(\nodes,p)$ random graphs (see Section \ref{sec:theory}), we used the empirical density of each real-world network, calculated from the numbers of nodes, $\nodes$, and edges, $m$, in each real-world graph as $p=m/{\nodes\choose 2}$. Our results are shown in Fig.\,\ref{fig:Fig4}.

In Fig.\,\ref{fig:Fig4}\,(a), we show the MSEs in using $\Srec$ to predict $\Savg$ for uniform-at-random node removal, plotted as a heatmap on a Voronoi diagram in $(\nodes,p)$ for the set of small real-world networks we used to test our model. The points corresponding to individual real-world networks around which the Voronoi cells are built are shown explicitly in the diagram for the larger cells. The errors observed in Fig.\,\ref{fig:Fig4} are typically very small and generally follow a similar trend to the errors observed in Fig.\,\ref{fig:Fig3}. 

Figure~\ref{fig:Fig4}\,(b) similarly presents results for node removal targeted by degree on the same set of small real-world networks. The errors in using $\Srec$ to predict $\Savg$ are noticeably worse for degree-targeted node removal than in the uniform-at-random removal case, but most MSE here are still below $0.01$. In Fig.\,\ref{fig:Fig4}\,(c), we show the MSE values for uniform-at-random and degree-targeted node removal as histograms, again emphasizing that overall the errors in using the $\Srec$ predictions are small, especially for uniform-at-random removal.

Three networks had extremely high MSEs for random and targeted node removals. We attribute these high MSEs to numerical issues due to the networks having very few edges and a large numbers of nodes\footnote{Their MSEs are $1.78\cdot 10^{21}$, $1.01e\cdot 10^{14}$, $1.68\cdot 10^{25}$, which correspond to $N$ values of 74, 70, and 78, and $p$ values of $0.0026$, $0.004$, and $0.003$ respectively.}, which result in  mean degrees $\langle k \rangle < 1$. We exclude these networks from the visualizations in Fig.\,\ref{fig:Fig4}. 

\section{Conclusion} \label{conclusion}
In this paper, we derived a model for accurately predicting the expected relative size of the largest connected component (LCC), $\langle S \rangle$, for small random graphs. We have shown that our model accurately captures the observed empirical average of the relative LCC size, $\Savg$, measured from simulated random graphs undergoing robustness tests through either uniform-at-random or degree-targeted node removals. In particular, our model is much more accurate for small networks than that obtained by application of the existing large-graph-limit theory. We also demonstrated the low errors that result from applying our method to measure the robustness of small real-world networks.
Because the LCC is an important performance measure that signals a network's robustness, with a higher LCC usually indicating higher network robustness, our model provides a tool to better measure and potentially improve real-world networks with various applications, for example by accurately comparing the empirical robustness of a given network to that of the corresponding random graph.

To build upon this model, future work might aim to improve the methods used here for estimating $\langle k_{max} \rangle$ in the predictions for degree-targeted node removals. Another possibly fruitful avenue of exploration might extend our method to predict $\langle S \rangle$ for random graphs undergoing other types of node removals, such as removing the node with highest betweenness centrality or damage \cite{lordan2019exact}.
To facilitate such work, we have included all of our codes at \cite{code}.

\begin{acknowledgments}
J.J. was supported by the Women in Science Program, Undergraduate Research Scholars at Dartmouth, and Kaminsky Family Fund. P.J.M. and A.S. were supported in part by the Army Research Office under MURI award W911NF-18-1-0244 and the National Science Foundation under award 2140024. The content is solely the responsibility of the authors and does not necessarily represent the official views of any agency supporting this research.
\end{acknowledgments}

\bibliography{references}

\providecommand{\noopsort}[1]{}\providecommand{\singleletter}[1]{#1}%
\begin{thebibliography}{12}%
\makeatletter
\providecommand \@ifxundefined [1]{%
 \@ifx{#1\undefined}
}%
\providecommand \@ifnum [1]{%
 \ifnum #1\expandafter \@firstoftwo
 \else \expandafter \@secondoftwo
 \fi
}%
\providecommand \@ifx [1]{%
 \ifx #1\expandafter \@firstoftwo
 \else \expandafter \@secondoftwo
 \fi
}%
\providecommand \natexlab [1]{#1}%
\providecommand \enquote  [1]{``#1''}%
\providecommand \bibnamefont  [1]{#1}%
\providecommand \bibfnamefont [1]{#1}%
\providecommand \citenamefont [1]{#1}%
\providecommand \href@noop [0]{\@secondoftwo}%
\providecommand \href [0]{\begingroup \@sanitize@url \@href}%
\providecommand \@href[1]{\@@startlink{#1}\@@href}%
\providecommand \@@href[1]{\endgroup#1\@@endlink}%
\providecommand \@sanitize@url [0]{\catcode `\\12\catcode `\$12\catcode
  `\&12\catcode `\#12\catcode `\^12\catcode `\_12\catcode `\%12\relax}%
\providecommand \@@startlink[1]{}%
\providecommand \@@endlink[0]{}%
\providecommand \url  [0]{\begingroup\@sanitize@url \@url }%
\providecommand \@url [1]{\endgroup\@href {#1}{\urlprefix }}%
\providecommand \urlprefix  [0]{URL }%
\providecommand \Eprint [0]{\href }%
\providecommand \doibase [0]{https://doi.org/}%
\providecommand \selectlanguage [0]{\@gobble}%
\providecommand \bibinfo  [0]{\@secondoftwo}%
\providecommand \bibfield  [0]{\@secondoftwo}%
\providecommand \translation [1]{[#1]}%
\providecommand \BibitemOpen [0]{}%
\providecommand \bibitemStop [0]{}%
\providecommand \bibitemNoStop [0]{.\EOS\space}%
\providecommand \EOS [0]{\spacefactor3000\relax}%
\providecommand \BibitemShut  [1]{\csname bibitem#1\endcsname}%
\let\auto@bib@innerbib\@empty
\bibitem [{\citenamefont {Holme}\ \emph {et~al.}(2002)\citenamefont {Holme},
  \citenamefont {Kim}, \citenamefont {Yoon},\ and\ \citenamefont
  {Han}}]{Holme2002}%
  \BibitemOpen
  \bibfield  {author} {\bibinfo {author} {\bibfnamefont {P.}~\bibnamefont
  {Holme}}, \bibinfo {author} {\bibfnamefont {B.~J.}\ \bibnamefont {Kim}},
  \bibinfo {author} {\bibfnamefont {C.~N.}\ \bibnamefont {Yoon}},\ and\
  \bibinfo {author} {\bibfnamefont {S.~K.}\ \bibnamefont {Han}},\ }\bibfield
  {title} {\bibinfo {title} {{Attack Vulnerability of Complex Networks}},\
  }\href {https://doi.org/10.1103/PhysRevE.65.056109} {\bibfield  {journal}
  {\bibinfo  {journal} {{Physical Review E}}\ }\textbf {\bibinfo {volume}
  {65}},\ \bibinfo {pages} {056109} (\bibinfo {year} {2002})}\BibitemShut
  {NoStop}%
\bibitem [{\citenamefont {Motter}(2004)}]{Motter2004}%
  \BibitemOpen
  \bibfield  {author} {\bibinfo {author} {\bibfnamefont {A.~E.}\ \bibnamefont
  {Motter}},\ }\bibfield  {title} {\bibinfo {title} {{Cascade Control and
  Defense in Complex Networks}},\ }\href
  {https://doi.org/10.1103/PhysRevLett.93.098701} {\bibfield  {journal}
  {\bibinfo  {journal} {{Physical Review Letters}}\ }\textbf {\bibinfo {volume}
  {93}},\ \bibinfo {pages} {098701} (\bibinfo {year} {2004})}\BibitemShut
  {NoStop}%
\bibitem [{\citenamefont {Louzada}\ \emph {et~al.}(2013)\citenamefont
  {Louzada}, \citenamefont {Daolio}, \citenamefont {Herrmann},\ and\
  \citenamefont {Tomassini}}]{Louzada2013}%
  \BibitemOpen
  \bibfield  {author} {\bibinfo {author} {\bibfnamefont {V.~H.}\ \bibnamefont
  {Louzada}}, \bibinfo {author} {\bibfnamefont {F.}~\bibnamefont {Daolio}},
  \bibinfo {author} {\bibfnamefont {H.~J.}\ \bibnamefont {Herrmann}},\ and\
  \bibinfo {author} {\bibfnamefont {M.}~\bibnamefont {Tomassini}},\ }\bibfield
  {title} {\bibinfo {title} {{Smart Rewiring for Network Robustness}},\ }\href
  {https://doi.org/10.1093/comnet/cnt010} {\bibfield  {journal} {\bibinfo
  {journal} {{Journal of Complex Networks}}\ }\textbf {\bibinfo {volume} {1}},\
  \bibinfo {pages} {150} (\bibinfo {year} {2013})}\BibitemShut {NoStop}%
\bibitem [{\citenamefont {Newman}(2018)}]{Newman}%
  \BibitemOpen
  \bibfield  {author} {\bibinfo {author} {\bibfnamefont {M.}~\bibnamefont
  {Newman}},\ }\href@noop {} {\emph {\bibinfo {title} {Networks}}}\ (\bibinfo
  {publisher} {Oxford University Press},\ \bibinfo {year} {2018})\BibitemShut
  {NoStop}%
\bibitem [{\citenamefont {Oehlers}\ and\ \citenamefont
  {Fabian}(2021)}]{Oehlers2021}%
  \BibitemOpen
  \bibfield  {author} {\bibinfo {author} {\bibfnamefont {M.}~\bibnamefont
  {Oehlers}}\ and\ \bibinfo {author} {\bibfnamefont {B.}~\bibnamefont
  {Fabian}},\ }\bibfield  {title} {\bibinfo {title} {{Graph metrics for network
  robustness---A survey}},\ }\href@noop {} {\bibfield  {journal} {\bibinfo
  {journal} {Mathematics}\ }\textbf {\bibinfo {volume} {9}},\ \bibinfo {pages}
  {895} (\bibinfo {year} {2021})}\BibitemShut {NoStop}%
\bibitem [{\citenamefont {Jiang}\ and\ \citenamefont {C}(2025)}]{code}%
  \BibitemOpen
  \bibfield  {author} {\bibinfo {author} {\bibfnamefont {J.}~\bibnamefont
  {Jiang}}\ and\ \bibinfo {author} {\bibfnamefont {S.~A.}\ \bibnamefont {C}},\
  }\href {https://github.com/acuschwarze/robustness-small-networks} {\bibinfo
  {title} {Code repository for `{R}obustness of small networks'}} (\bibinfo
  {year} {2025})\BibitemShut {NoStop}%
\bibitem [{\citenamefont {Sahimi}(2023)}]{Sahimi2023}%
  \BibitemOpen
  \bibfield  {author} {\bibinfo {author} {\bibfnamefont {M.}~\bibnamefont
  {Sahimi}},\ }\href@noop {} {\emph {\bibinfo {title} {{Applications of
  Percolation Theory}}}},\ \bibinfo {edition} {2nd}\ ed.\ (\bibinfo
  {publisher} {Springer},\ \bibinfo {address} {{Cham, Switzerland}},\ \bibinfo
  {year} {2023})\BibitemShut {NoStop}%
\bibitem [{\citenamefont {Schwarze}\ \emph {et~al.}(2024)\citenamefont
  {Schwarze}, \citenamefont {Jiang}, \citenamefont {Wray},\ and\ \citenamefont
  {Porter}}]{schwarze2024}%
  \BibitemOpen
  \bibfield  {author} {\bibinfo {author} {\bibfnamefont {A.~C.}\ \bibnamefont
  {Schwarze}}, \bibinfo {author} {\bibfnamefont {J.}~\bibnamefont {Jiang}},
  \bibinfo {author} {\bibfnamefont {J.}~\bibnamefont {Wray}},\ and\ \bibinfo
  {author} {\bibfnamefont {M.~A.}\ \bibnamefont {Porter}},\ }\bibfield  {title}
  {\bibinfo {title} {Structural robustness and vulnerability of networks},\
  }\href@noop {} {\bibfield  {journal} {\bibinfo  {journal} {arXiv preprint
  arXiv:2409.07498}\ } (\bibinfo {year} {2024})}\BibitemShut {NoStop}%
\bibitem [{\citenamefont {Albert}\ \emph {et~al.}(2000)\citenamefont {Albert},
  \citenamefont {Jeong},\ and\ \citenamefont
  {Barab{\'a}si}}]{Barabasi-Jeong-Albert}%
  \BibitemOpen
  \bibfield  {author} {\bibinfo {author} {\bibfnamefont {R.}~\bibnamefont
  {Albert}}, \bibinfo {author} {\bibfnamefont {H.}~\bibnamefont {Jeong}},\ and\
  \bibinfo {author} {\bibfnamefont {A.-L.}\ \bibnamefont {Barab{\'a}si}},\
  }\bibfield  {title} {\bibinfo {title} {Error and attack tolerance of complex
  networks},\ }\href@noop {} {\bibfield  {journal} {\bibinfo  {journal}
  {Nature}\ }\textbf {\bibinfo {volume} {406}},\ \bibinfo {pages} {378}
  (\bibinfo {year} {2000})}\BibitemShut {NoStop}%
\bibitem [{\citenamefont {van~der {H}ofstad}(2017)}]{vanDerHofstadt2017}%
  \BibitemOpen
  \bibfield  {author} {\bibinfo {author} {\bibfnamefont {R.}~\bibnamefont
  {van~der {H}ofstad}},\ }\href@noop {} {\emph {\bibinfo {title} {{Random
  Graphs and Complex Networks}}}},\ Vol.~\bibinfo {volume} {1}\ (\bibinfo
  {publisher} {{Cambridge University Press}},\ \bibinfo {address} {{Cambridge,
  United Kingdom}},\ \bibinfo {year} {2017})\BibitemShut {NoStop}%
\bibitem [{Note1()}]{Note1}%
  \BibitemOpen
  \bibinfo {note} {Their MSEs are $1.78\cdot 10^{21}$, $1.01e\cdot 10^{14}$,
  $1.68\cdot 10^{25}$, which correspond to $N$ values of 74, 70, and 78, and
  $p$ values of $0.0026$, $0.004$, and $0.003$ respectively.}\BibitemShut
  {Stop}%
\bibitem [{\citenamefont {Lordan}\ and\ \citenamefont
  {Albareda-Sambola}(2019)}]{lordan2019exact}%
  \BibitemOpen
  \bibfield  {author} {\bibinfo {author} {\bibfnamefont {O.}~\bibnamefont
  {Lordan}}\ and\ \bibinfo {author} {\bibfnamefont {M.}~\bibnamefont
  {Albareda-Sambola}},\ }\bibfield  {title} {\bibinfo {title} {Exact
  calculation of network robustness},\ }\href@noop {} {\bibfield  {journal}
  {\bibinfo  {journal} {Reliability Engineering \& System Safety}\ }\textbf
  {\bibinfo {volume} {183}},\ \bibinfo {pages} {276} (\bibinfo {year}
  {2019})}\BibitemShut {NoStop}%
\end{thebibliography}%

\end{document}